\documentclass[twocolumn,longauthor]{aastex62}
\usepackage{amsmath}
\usepackage{comment}
\usepackage{xcolor}
\usepackage{hyperref}
\usepackage{anyfontsize}

\definecolor{reddish}{rgb}{0.813642,0.212345,0.196948}
\definecolor{blueish}{rgb}{0.0745098,0.52549,0.776471}
		
\hypersetup{
        colorlinks=true,        % false: boxed links; true: colored links
        urlcolor=blueish,          % color of external links
        pdfpagelabels=true,
        hypertexnames=true,
        plainpages=false,
        naturalnames=true,
        pdftitle={The Halo Drive},    % title
        pdfauthor={Kipping},     % author
        linkcolor=reddish,          % color of internal links (change box color with linkbordercolor)
        citecolor=blueish,        % color of links to bibliography
}

\shorttitle{The Halo Drive}
\shortauthors{Kipping}
\begin{document}

\title{THE HALO DRIVE:\\
FUEL-FREE RELATIVISTIC PROPULSION OF LARGE MASSES VIA RECYCLED BOOMERANG PHOTONS}

\correspondingauthor{David Kipping}
\email{dkipping@astro.columbia.edu}

\author[0000-0002-4365-7366]{{\fontsize{10.5}{12.6}\selectfont \textcolor{black}{David Kipping$^1$}}}
\affil{$^1$Department of Astronomy,
Columbia University,
550 W 120th Street,
New York, NY 10027, USA}

% Accepted for publication 2nd January 2019

%% Note that the \and command from previous versions of AASTeX is now
%% depreciated in this version as it is no longer necessary. AASTeX 
%% automatically takes care of all commas and "and"s between authors names.

%% AASTeX 6.2 has the new \collaboration and \nocollaboration commands to
%% provide the collaboration status of a group of authors. These commands 
%% can be used either before or after the list of corresponding authors. The
%% argument for \collaboration is the collaboration identifier. Authors are
%% encouraged to surround collaboration identifiers with ()s. The 
%% \nocollaboration command takes no argument and exists to indicate that
%% the nearby authors are not part of surrounding collaborations.

%% Mark off the abstract in the ``abstract'' environment. 
\begin{abstract}
Gravitational slingshots around a neutron star in a compact binary have been
proposed as a means of accelerating large masses to potentially relativistic
speeds. Such a slingshot is attractive since fuel is not expended for the
acceleration, however it does entail a spacecraft diving into close proximity of
the binary, which could be hazardous. It is proposed here that such a slingshot can
be performed remotely using a beam of light which follows a boomerang null
geodesic. Using a moving black hole as a gravitational mirror, kinetic energy from the
black hole is transferred to the beam of light as a blueshift and upon return
the recycled photons not only accelerate, but also add energy to, the
spacecraft. It is shown here that this gained energy can be later expended to
reach a terminal velocity of approximately 133\% the velocity of the black
hole. A civilization could exploit black holes as galactic way points but
would be difficult to detect remotely, except for an elevated binary merger
rate and excess binary eccentricity.
\end{abstract}


\section{Introduction}
\label{sec:intro}

In recent months, there has been increased interest in light sailing propulsion
systems, including using direct energy, thanks (in part) to the
\textit{Breakthrough Starshot} project announced in 2016. Since the early
$20^{\mathrm{th}}$ century, it has been recognized that the momentum carried
by light could be used to accelerate spacecraft \citep{zander:1925}. Although
the momentum exchanges are tiny, what makes radiation pressure attractive as
a propulsion system is the fact that fuel need not be carried by the spacecraft
itself. Either through Solar radiation \citep{garwin:1958} or directed energy
\citep{marx:1966,redding:1967,forward:1984}, such systems could
be used to overcome the limitations imposed by the Tsiolkovsky
rocket equation affecting conventional reaction drives.

Achieving relativistic speeds through such a system is theoretically achievable
by directing high powered lasers at spacecraft (see \citealt{bible:2013,
benford:2013}). For non-relativistic speeds, the energy required to accelerate
a spacecraft of mass $m$ to velocity $\beta c$ equals $\tfrac{\beta}{2} m c^2$
(via a first-order expansion in $\beta$ of Equation~(6) of
\citealt{kulkarni:2016}). This highlights that accelerating massive objects
to relativistic speeds is certainly challenging since one needs to first store
energy comparable to the rest mass. Accelerating a low-mass ($\sim$ gram)
spacecraft may be feasible albeit at considerable energetic cost
($\sim10$\,TJ), but larger masses pose severe challenges.

An idealized propulsion system would be able to accelerate arbitrarily
large masses to relativistic speeds at little to no energy cost. At first, this
statement may seem fanciful yet essentially free speed-boosts have been
exploited for decades in the Solar System via gravitational assists,
although not to the speeds associated with relativistic flight.
Perhaps the ultimate incarnation of the gravity assist was proposed by
\citet{dyson:1963}, who argued that a compact binary of white dwarfs
or neutron stars could be exploited to accelerate arbitrarily large masses up
to relativistic speeds (assuming the binary is sufficiently compact).
This ``Dyson slingshot'' maneuver is theoretically attractive but
swinging round a neutron star in close proximity is potentially hazardous
due to extreme tidal forces and the circumbinary radiation environment.

In this work, it is shown that the Dyson slingshot can be performed remotely
using the ideas from directed energy light sailing and gravitational mirrors.
Gravitational mirrors were first described in \citet{stuckey:1993} who
showed that null geodesics exists around Schwarzschild black holes enabling
one to see one's own reflection. Photons just skimming the
photon sphere perform a full revolution and can make their way back to
the source, dubbed as ``boomerang photons'' by the author. Using this effect,
it is argued here that a moving black hole can be used like a moving mirror,
causing light to not only return to the source but also receive a blue
shift due to the black hole's relative motion. Photons are recycled by the
spacecraft and repeatedly emitted and re-absorbed from the gravitational
mirror, accelerating the spacecraft up to speeds ultimately exceeding that
of the black hole itself.

For convenience, this setup is referred as a ``halo drive'' in what
follows, as a result of the ring of light which wraps around the black
hole and the propulsive nature of the final outcome. Such a system is
capable of achieving the Dyson slingshot but without requiring a spacecraft
to become in close proximity of the binary itself. Whilst slingshots
could be performed around an isolated moving black hole, binaries are
focused on in this work due to their potential for compact configurations
where relativistic speeds could be achieved (although the expressions
derived throughout are equally applicable to isolated black holes too).
With $\mathcal{O}[10^8]$ black holes estimated to reside within the Milky
Way \citep{elbert:2017}, a large network of way-points potentially
exist to permit intra-galactic travel. This work describes some of the
basic mathematics behind the halo drive concept and the consequences for both
the spacecraft and the binary itself.

\section{Deflection off a Moving Black Hole}
\label{sec:deflection}

\subsection{A halo from boomerang photons}
\label{sub:geodesic}

The majority of this paper will concern itself will computing the velocities
which can be achieved by a spacecraft using the halo drive described in
Section~\ref{sec:intro}. However, it is worth first establishing that
boomerang photon geodesics exist and considering the shape of such paths.

Boomerang null geodesics were first introduced by \citet{stuckey:1993}, who
considered the Schwarzschild metric and demonstrated that such geodesics
exist and effectively turn black holes in gravitational mirrors. Indeed,
theoretically an infinite number of distinct boomerang geodesics exist,
corresponding to how many loops around the black hole are conducted. If
one writes the critical impact parameter for photon capture as $b_c$, then
the number of loops of a scattered photon equals $N \sim -\log(-1+b/b_c)/2\pi$
\citep{zeldovich:1971}. However, for the sake of this work, the scope is
limited to that of the first-order geodesic which does not perform multiple
revolutions.

\begin{figure}
\begin{center}
\includegraphics[width=8.4cm,angle=0,clip=true]{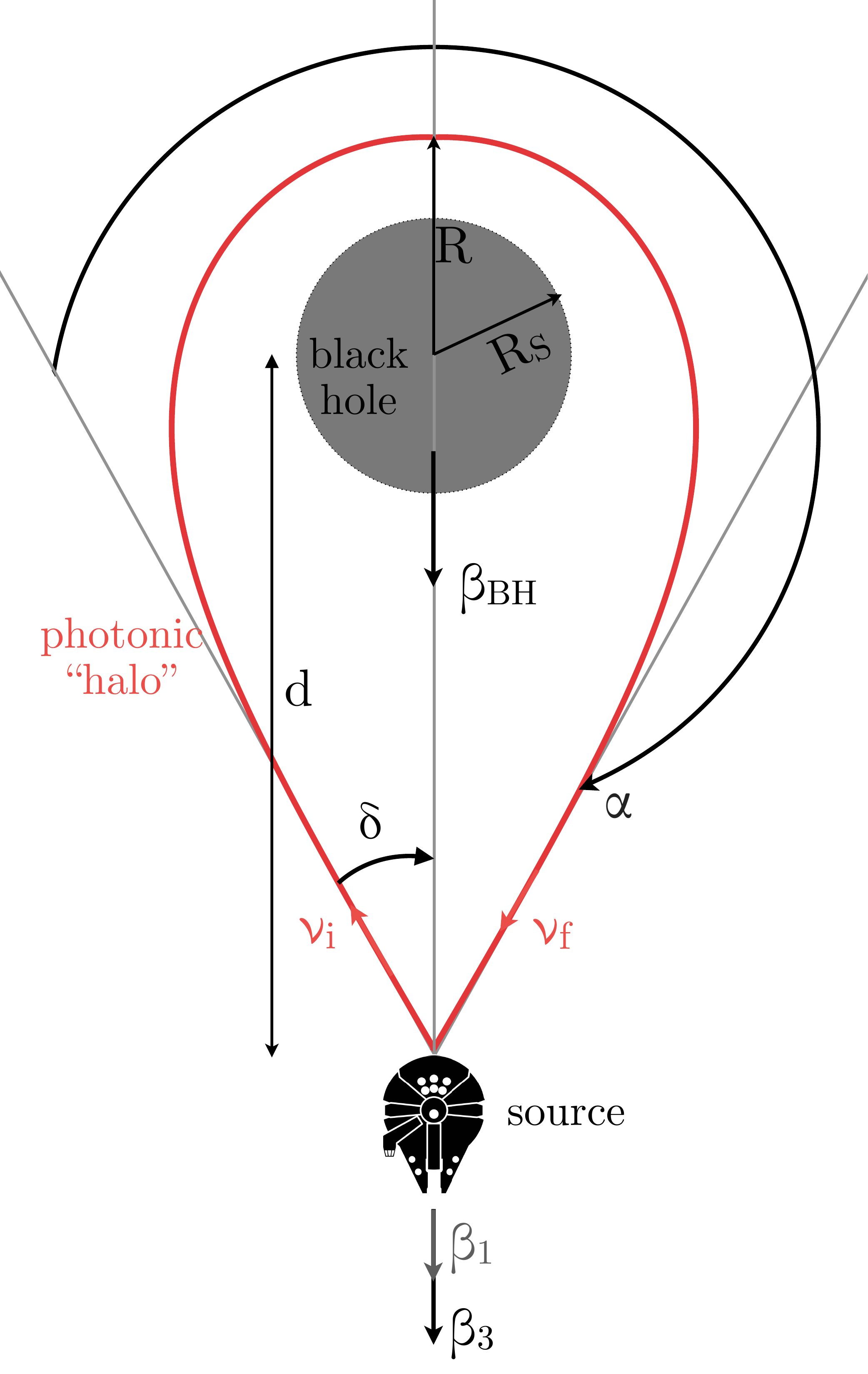}
\caption{
Outline of the halo drive.
A spaceship traveling at a velocity $\beta_i$ emits a photon of frequency
$\nu_i$ at a specific angle $\delta$ such that the photon completes a halo
around the black hole, returning shifted to $\nu_f$ due to the forward
motion of the black hole, $\beta_{\mathrm{BH}}$.
}
\label{fig:halocartoon}
\end{center}
\end{figure}

Light is emitted from the source at an angle $\delta$ relative to the
radial direction and experiences strong deflection as it approaches the event
horizon. For a Schwarzschild boomerang geodesic, there is rotational symmetry
about the radial direction meaning that the angle of emission equals the
angle of reception \citep{stuckey:1993}. The basic setup is depicted in
Figure~\ref{fig:halocartoon}.

In order for the deflection to be strong enough to constitute a boomerang, this
requires the light's closest approach to the black hole to be within a couple
of Schwarzschild radii, $R_S \equiv 2 G M/c^2$. Light which makes a closest
approach smaller than $3 G M/c^2$ becomes trapped in orbit, known as the photon
sphere, and thus typical boomerang geodesics skim just above this critical
distance.

\begin{figure*}
\begin{center}
\includegraphics[width=17.0cm,angle=0,clip=true]{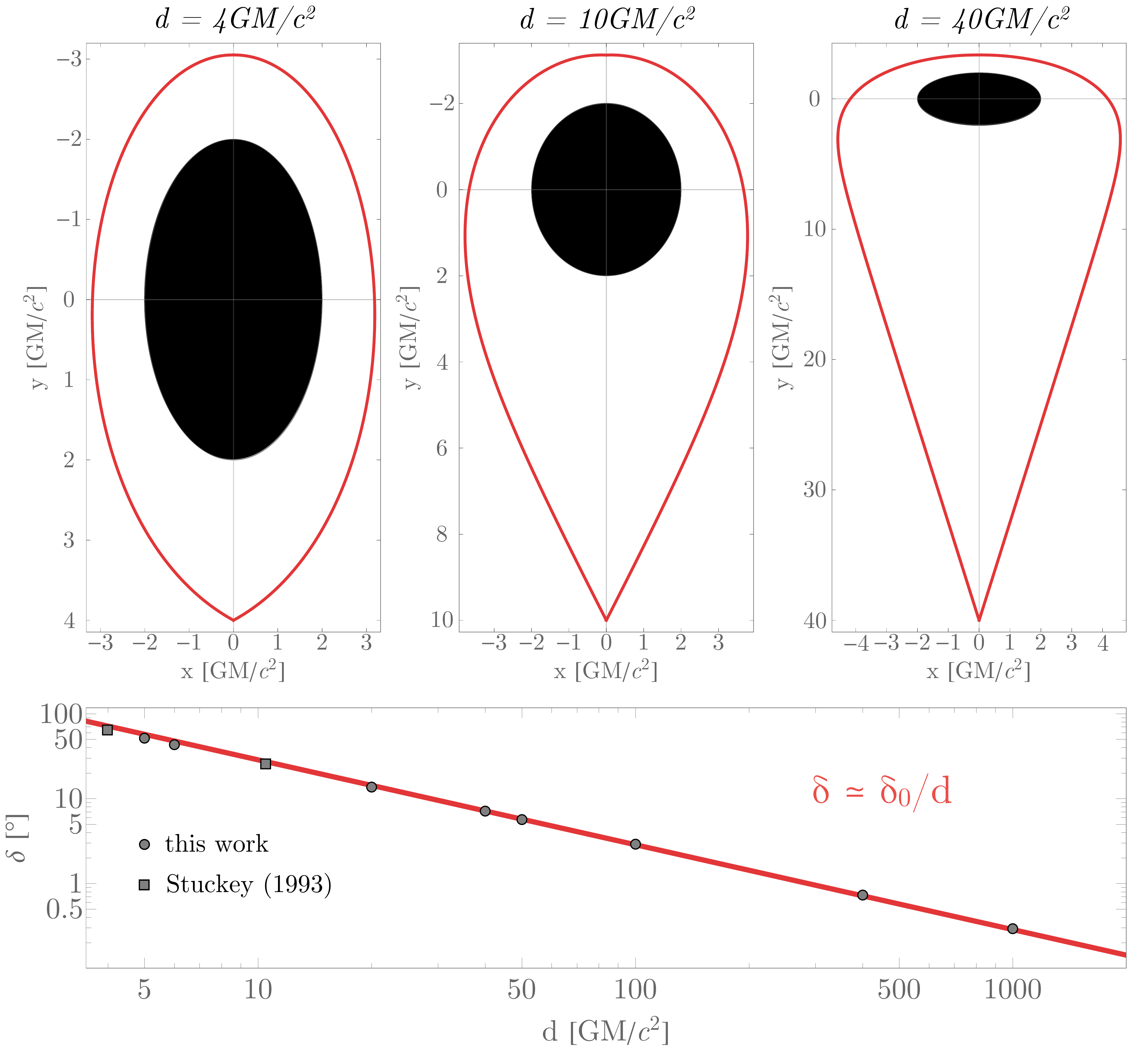}
\caption{
Top panel shows three numerically integrated boomerang null geodesics where
the initial standoff distance, $d$, is varied. Solving for the boomerang
deflection angle for a series of different $d$ values, it may be seen that
the angle drops off as $\propto 1/d$ to a good approximation, especially
when $d \gg G M/c^2$.
}
\label{fig:haloexamples}
\end{center}
\end{figure*}

\citet{stuckey:1993} showed that boomerang null geodesics could be computed by
numerically integrating the rate of change of the radial coordinate with
respect to the azimuthal coordinate, $\mathrm{d}r/\mathrm{d}\varphi$ (a simple
algorithm is described in the Appendix of that work). To illustrate this,
numerical integrations of the geodesic were performed with $10^6$ steps for a
series of different initial standoff distances, $d$. As shown in
Figure~\ref{fig:haloexamples}, the critical
deflection necessary to perform a boomerang appears to be proportional to
$1/d$ to a good approximation (particularly when $d \gg G M/c^2$), with a
constant of proportionality given by $\delta_0 = 286.5^{\circ}$.

More importantly for this work, the experiment described above demonstrates
that $\delta \to 0$ as $d$ becomes large and thus the angle $\alpha$
depicted in Figure~\ref{fig:halocartoon} approaches $\pi$ radians
(since $\alpha = \pi - 2\delta$). This simplification will be exploited
this later in Section~\ref{sec:engine}.

Even in the idealized Schwarzschild case, the results shown above are
not generally applicable to a practical halo drive. This is because
the photon should not return to the precise same location but rather
a greater radial distance, since the spacecraft will experience a
back-reaction after emission (or exhaust) of the photon. Thus, the
angle should be chosen according to the rate of acceleration desired.

For the sake of demonstrating the principle of the halo drive, this
paper does not concern itself with the precise angular correction
needed to accomplish this. It is worth highlighting that boomerang geodesics
can be constructed in the more general case of a Kerr metric, as
discussed in \citet{cramer:1997}. Since the halo drive exploits a
compact binary, both components should be included in a more precise
calculation. For the sake of this work, it is sufficient to note that
such a) geodesics exist and are computable b) the angle
$\alpha \simeq \pi$ when $d \gg G M/c^2$, simplifying subsequent
calculations. Note that point a) could be calculated onboard
the spacecraft either using a metric known to be completely correct,
or using a pilot low-power laser to fine-tune the correct angle.

\subsection{Deflections in the black hole's rest frame}
\label{sub:deflections1}

Let us now turn to calculating the movement of a spacecraft in response to
emitting a boomerang photon (or halo) around a moving black hole. Before
considering the effect on the spacecraft, one needs to first derive the changes
imparted onto a photon which conducts such a loop. 

\begin{figure}
\begin{center}
\includegraphics[width=8.4cm,angle=0,clip=true]{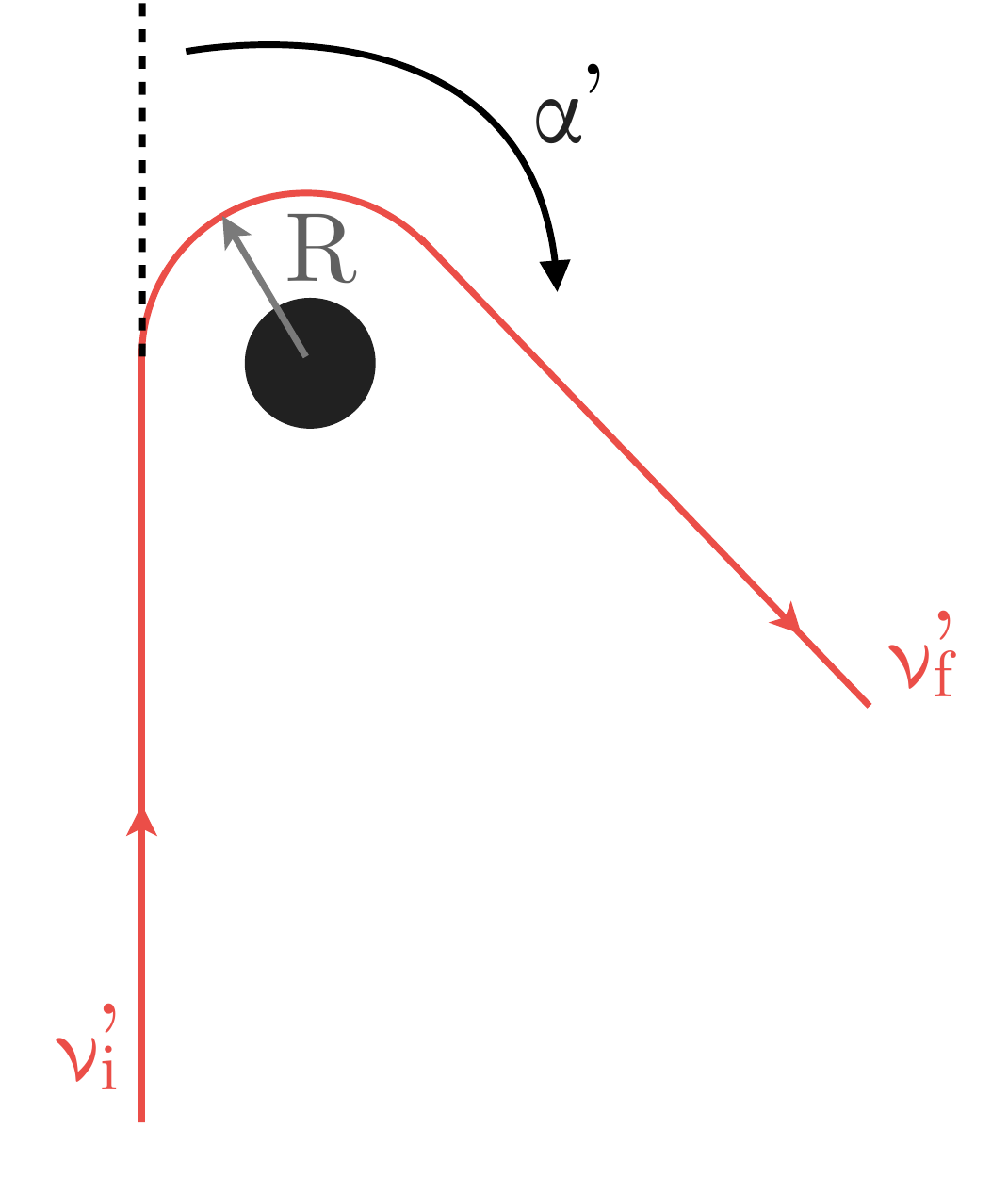}
\caption{
An incident photon of frequency $\nu_i'$ is deflected around a
Schwarzschild black hole by an angle $\alpha'$, depicted here
in the rest frame of the BH.
}
\label{fig:cartoon}
\end{center}
\end{figure}

Let us work in the rest frame of the black hole and assuming that
an incident photon passes by with an impact parameter exceeding $3\sqrt{3}
G M/c^2$, such that the closest approach exceeds $3 G M/c^2$. In such a
case, it is expected that the photon to be deflected by some arbitrary angle (see
\citealt{darwin:1959}) which is labelled as $\alpha'$, as depicted in
Figure~\ref{fig:cartoon}.

As described in Section~\ref{sub:geodesic}, the angle $\alpha'$ is set by the
mass of the BH and the impact parameter of the encounter. To start, let us consider
what the frequency of the deflected photon, $\nu_f'$, will be. This can be
computed by conserving relativistic four-momentum before and after the
encounter. The initial four-momentum of the photon, working in units of $c$,
is given by

\begin{equation}
\mathbf{P}_{\mathrm{EM},\mathrm{pre}} = \begin{pmatrix}
h \nu_i' \cr
0 \cr
h \nu_i' \cr
0
\end{pmatrix},
\end{equation}

and that of the black hole of

\begin{equation}
\mathbf{P}_{\mathrm{BH},\mathrm{pre}} = \begin{pmatrix}
M \cr
0 \cr
0 \cr
0
\end{pmatrix}.
\end{equation}

Let us write the final four-momentum vectors of these components as

\begin{equation}
\mathbf{P}_{\mathrm{EM},\mathrm{post}} = \begin{pmatrix}
h \nu_f' \cr
h \nu_f' \sin \alpha' \cr
h \nu_f' \cos \alpha' \cr
0
\end{pmatrix},
\end{equation}

and

\begin{equation}
\mathbf{P}_{\mathrm{BH},\mathrm{post}} = \begin{pmatrix}
E \cr
p \sin \theta \cr
p \cos \theta \cr
0
\end{pmatrix}.
\end{equation}

Conserving each component of the total four-momentum, one finds

\begin{align}
h \nu_i' + M &= h \nu_f' + E,\\
0 &= h \nu_f' \sin \alpha' + p \sin \theta,\\
h \nu_i' &= h \nu_f' \cos \alpha' + p \cos \theta.
\end{align}

Taking the last two lines, then squaring and summing, one may write

\begin{align}
p^2 &= (h \nu_f' \sin \alpha')^2 + (h \nu_i' - h \nu_f' \cos \alpha')^2.
\end{align}

One may now use the relation $E^2 = p^2 + M^2$ and our earlier energy
expression to write that

\begin{align}
(h \nu_i' + M - h \nu_f')^2 &= (h \nu_f' \sin \alpha')^2 + (h \nu_i' - h \nu_f' \cos \alpha')^2 + M^2.
\end{align}

Solving the above for $\nu_f'$ yields the familiar Compton scattering equation

\begin{align}
\nu_f' &= \frac{ \nu_i' }{ 1 + \tfrac{h \nu_i'}{M c^2} (1-\cos\alpha')},
\label{eqn:compton}
\end{align}

where the $c$ units have been re-added.

\subsection{Deflections around a moving black hole}
\label{sub:deflections2}

The general principle of the halo drive is to siphon kinetic energy
from the BH and thus one ultimately requires computing deflections
around a moving BH. Armed with the result from the previous subsection,
this can be easily accomplished using Lorentz transforms.

Let us assume that the black hole is moving along the $-\hat{y}$ direction
such that its velocity vector is given by $\mathbf{v} =
\{0,-\beta_{\mathrm{BH}} c,0\}^T$. In such a case, the incident photon's
energy in the observer's frame, $\nu_i$, may be related to that in the
black's hole (initial) rest frame, $\nu_i'$, using

\begin{align}
\nu_i' &= \nu_i \sqrt{\frac{ 1 + \beta_{\mathrm{BH}} }{ 1 - \beta_{\mathrm{BH}} }}.
\end{align}

The deflected photon returns at an arbitrary angle and so using the
more general boosting expression yields

\begin{align}
\nu_f &= \nu_f' \frac{ \sqrt{1-\beta_{\mathrm{BH}}^2} }{ 1 + \beta_{\mathrm{BH}} \cos \alpha }.
\end{align}

One may now account for the photon's deflection by substituting
$\nu_f'$ using Equation~(\ref{eqn:compton}) to obtain

\begin{align}
\nu_f &= \nu_i \frac{ \sqrt{\frac{ 1 + \beta_{\mathrm{BH}} }{ 1 - \beta_{\mathrm{BH}} }} }{ 1 + \sqrt{\frac{ 1 + \beta_{\mathrm{BH}} }{ 1 - \beta_{\mathrm{BH}} }} \tfrac{h \nu_i}{M c^2} (1-\cos\alpha')} \frac{ \sqrt{1-\beta_{\mathrm{BH}}^2} }{ 1 + \beta_{\mathrm{BH}} \cos \alpha }.
\label{eqn:nufunpolished}
\end{align}

It is worth highlighting that in the limit of $M\to\infty$ (an infinite mass
mirror) and $\alpha \to \pi$ (a normal reflection),
Equation~(\ref{eqn:nufunpolished}) reproduces same familiar result as that of
\citep{einstein:1905}, i.e.

\begin{align}
\lim_{M\to\infty} {\alpha \to \pi} \nu_f &= \nu_i \Big(\frac{ 1 + \beta_{\mathrm{BH}} }{ 1 - \beta_{\mathrm{BH}} }\Big).
\end{align}

Finally, one needs to relate $\alpha'$, the angle of deflection in the
black hole's initial rest frame, to $\alpha$, the angle of deflection in
the observer's frame. Consider the setup as depicted in
Figure~\ref{fig:cartoon} where $\alpha'$ is obtuse. If one defines an angle
$\theta'=\pi-\alpha$ as the acute and opposite angle, this angle along the
direction of motion should be expected to be squeezed in the moving frame
as a result of relativistic aberation. Accordingly, one would expect the
observer's frame to have $\theta'>\theta$, or more explicitly using the
aberation formula

\begin{align}
\cos \theta &= \frac{ \cos \theta' - \beta_{\mathrm{BH}} }{ 1 - \beta_{\mathrm{BH}} \cos \theta'}.
\end{align}

Re-writing in terms of $\alpha$ angles and re-arranging to make $\alpha'$
the subject, one finds

\begin{align}
\cos \alpha' &= \frac{ \beta_{\mathrm{BH}} - \cos \alpha }{ \beta_{\mathrm{BH}} \cos \alpha - 1 }.
\end{align}

Plugging this into our earlier result for the frequency shift
given by Equation~(\ref{eqn:nufunpolished}), one may write

\begin{align}
\nu_f &= \nu_i \Bigg( \frac{ 1+\beta_{\mathrm{BH}} }{ 1 + \beta_{\mathrm{BH}} \cos \alpha } \Bigg) \Bigg(\frac{ 1 }{ 1 + K_i \sqrt{\frac{ 1 + \beta_{\mathrm{BH}} }{ 1 - \beta_{\mathrm{BH}} }} (1-\frac{ \beta_{\mathrm{BH}} - \cos \alpha }{ \beta_{\mathrm{BH}} \cos \alpha - 1 })} \Bigg),
\label{eqn:nuf}
\end{align}

where

\begin{align}
K_i &\equiv \frac{h \nu_i}{M c^2}.
\end{align}

In the limit where $h \nu_i \ll M c^2$ (the infinite mass limit), then the
photon's frequency upon return is well-approximated by

\begin{align}
\lim_{K_i \to 0} \nu_f &= \nu_i \Big(\frac{ 1 + \beta_{\mathrm{BH}} }{ 1 + \beta_{\mathrm{BH}} \cos \alpha }\Big).
\label{eqn:infmass}
\end{align}

\subsection{Gravitational redshifting during the deflection}
\label{sub:GRshift1}

One effect that has been ignored thus far is gravitational blue/red shift.
If the photon is assumed to return to the same location it originated
from, then the net change in gravitational potential energy from emission
to reflection is zero. However, during the approach of the photon, it will
become increasingly blue, potentially affecting our expressions.
It is argued here that this effect is extremely small and can be safely
ignored for the purposes of this paper, although could be accounted for
using numerical integrations.

Let us take the quite reasonable assumption that $M c^2 \gg h \nu_i$,
such that Equation~(\ref{eqn:nuf}) can be approximated to
Equation~(\ref{eqn:infmass}). Let us denote the radial separation of
the photon from the black hole as $d[t]$ i.e. as a function of time.
Therefore, during the approach, one expects the photon's frequency
to be blue shifted as

\begin{align}
\nu_i[t] &= \nu_i \sqrt{\frac{d[t] (d_0-R_S)}{d_0 (d[t]-R_S)}}.
\end{align}

In other words, it is simply a multiplicative factor of the original
frequency. The photon is then shifted by the deflection encounter
according to Equation~(\ref{eqn:infmass}), which is again
simply a multiplicative factor of frequency. Finally, upon
return the equal and opposite gravitational redshift occurs
(since the photon returns to the same location), cancelling
out the previous blue shift.

Accordingly, in the limit of $M c^2 \gg h \nu_i$ and the photon
returning to the same location, gravitational blue/red shift
has zero net effect. This symmetry is broken when one includes the
$K_i$ term, and again this could be correctly accounted for
using numerical integrations, however it is a fairly extreme
case that is technically forbidden as long as the spaceship
has a low mass compared to the BH i.e. $m \ll M$ (see
Section~\ref{sub:GRshift2} for justification).

If the photon does not return to the same location but at a
greater radial distance (as expected since the spacecraft
will be in motion), then there will be a net effect
even in the limit of $M c^2 \gg h \nu_i$. This is discussed
later in Section~\ref{sub:GRshift2}.

\section{Formalism for the Halo Drive}
\label{sec:engine}

\subsection{Response of a spacecraft}

Consider an initial setup where a spacecraft of mass $m_1$ resides in a wide
orbit around a binary BH. At one of the quadrature point in the binary orbit,
the BH will be approaching the spacecraft at a relative velocity of
$\beta_{\mathrm{BH}}$. More generally, the spacecraft may have already begun
to accelerate away from the black hole and thus have a velocity of $\beta_1$
in the same direction.

The source (or spacecraft) emits a photon of energy $\nu_i$ and this will lead
to a slight back impulse on the source. The source must also slightly decrease
in mass as a result of the emission, reducing from $m_1$ to $m_2$, culminating
in the source increasing in speed from $\beta_1$ to $\beta_2$. Conserving
relativistic energy and momentum, one may write that

\begin{align}
\frac{m_1 c^2}{\sqrt{ 1 - \beta_1^2 }} &= h \nu_i + \frac{m_2 c^2}{\sqrt{1-\beta_2^2}},\nonumber\\
\frac{m_1 c \beta_1}{\sqrt{1-\beta_1^2}} &= -\frac{h \nu_i}{c} + \frac{m_2 c \beta_2}{\sqrt{1-\beta_2^2}},
\end{align}

where $m$ is the mass of the source. Solving the above and simplifying,
one finds a speed of

\begin{align}
\beta_2 =& \frac{ \beta_1 + \kappa_{i1} \sqrt{1-\beta_1^2} }{ 1 - r_{i1} \sqrt{1-\beta_{i1}^2} },
\label{eqn:beta2}
\end{align}

and a mass of

\begin{align}
\frac{m_2}{m_1} &= \sqrt{ 1 - 2 \kappa_{i1} \sqrt{ \frac{1+\beta_1}{1-\beta_1}} },
\label{eqn:mass2}
\end{align}

where in both expressions

\begin{align}
\kappa_{i1} &\equiv \frac{ h \nu_i}{m_1 c^2}.
\end{align}

The spacecraft has finite mass and so cannot emit photons of arbitrary
energy. Taking the resulting equation for $m_2/m_1$, one may solve that the
limiting case is when the mass approaches zero, the maximum allowed
photon emission corresponds to

\begin{align}
\kappa_{i1,\mathrm{max}} &= \frac{1}{2} \sqrt{\frac{1-\beta_1}{1+\beta_1}}.
\label{eqn:ri1max}
\end{align}

In this where the mass is totally converted into energy, the final
speed of the now massless spacecraft can be shown to equal $c$, since
it is essentially just the returning photon.

The results can be combined with the change once the photon returns with
a modified frequency $\nu_f$. Consider the simplified case where the final and
initial position of the source are both sufficiently out of the gravitational
well that the effects of gravitational redshift can be ignored. Further, the
relativistic Doppler effect that occurs between the returning photon and the
now-moving spacecraft is ignored, such that $\nu_f$ is given
by Equation~(\ref{eqn:infmass}). The Doppler effect will be accounted for later
in Section~\ref{sub:doppler}. Under these assumptions, one can construct another
set of equations for the absorption given by

\begin{align}
\frac{m_2 c^2}{\sqrt{ 1 - \beta_2^2 }} + h \nu_f &= \frac{m_3 c^2}{\sqrt{1-\beta_3^2}},\nonumber\\
\frac{m_2 c \beta_2}{\sqrt{1-\beta_2^2}} + \frac{h \nu_f}{c} &= \frac{m_3 c \beta_3}{\sqrt{1-\beta_3^2}},
\end{align}

giving

\begin{align}
\beta_3 =& \frac{ \beta_2 + \kappa_{f2} \sqrt{1-\beta_2^2} }{ 1 + \kappa_{f2} \sqrt{1-\beta_2^2} },
\label{eqn:beta3}
\end{align}

and

\begin{align}
\frac{m_3}{m_2} =& \sqrt{ 1 + 2 \kappa_{f2} \sqrt{ \frac{1-\beta_2}{1+\beta_2} } },
\label{eqn:mass3}
\end{align}

where in both expressions

\begin{align}
\kappa_{f2} &\equiv \frac{ h \nu_f}{m_2 c^2}.
\end{align}

Note that the latter of the new equations reveals that $\lim_{m_2 \to 0} m_3
= 0$, which happens when $\kappa_{i1} \to \kappa_{i1,\mathrm{max}}$. In other
words, if the spacecraft converts all of its mass into energy and returns as a
pure photon, there is no mechanism here for the photon to somehow return back
to massive spacecraft. Substituting in the earlier equations, and after much
simplification, one finds that

\begin{align}
\lim_{K_i \to 0} \lim_{\alpha \to \pi} \beta_3 &= \frac{ \beta_1 (1 - \beta_{\mathrm{BH}}) + 2 \kappa_{i1} \sqrt{1-\beta_1^2} }{ (1 - \beta_{\mathrm{BH}}) + 2 \kappa_{i1} \beta_{\mathrm{BH}} \sqrt{1 - \beta_1^2}}.
\label{eqn:beta3simple}
\end{align}

In the limit of the photon's carrying no momentum ($\kappa_{i1} \to 0$), then
the final velocity is unchanged from the initial velocity, as expected. In the
limit of the intermediate mass, $m_2$, being zero implying a complete
conversion into energy, the final speed is $c$ as expected for a massless
particle. It is worth highlighting that in the limit of $\kappa_{i1} \to 0$, which is
to say the back-reaction effect described in \citet{kipping:2017} is ignored,
then $\beta_3 \to \beta_1$ and no acceleration is achieved, underlining the
importance of the effect described in that paper.

Although the velocity change in Equation~(\ref{eqn:beta3simple}) is small for
low choices of $\kappa_{i1}$, it is emphasized that any number of photons can be
fired and at any frequency and these velocity differences accumulate. At each
stage, not only is the source accelerated, but it is also gains mass (or energy).
Specifically, the mass change is given by

\begin{align}
\lim_{K_i \to 0 } \lim_{ \alpha \to \pi } \frac{m_3}{m_1} =& \sqrt{
1 - 2 \kappa_{i1} \sqrt{\frac{1+\beta_1}{1-\beta_1}} } \nonumber\\
\qquad& \sqrt{
1 + 2 \kappa_{i1} \sqrt{\frac{1-\beta_1}{1+\beta_1}} \Big(\frac{1+\beta_{\mathrm{BH}}}{1-\beta_{\mathrm{BH}}}\Big) }.
\label{eqn:masschange}
\end{align}

Note that $m_3$ equals $m_1$ if $\kappa_{i1} \to 0$, demonstrating again that if
the \citet{kipping:2017} back-reaction effect is ignored, the mass of the
spacecraft would be unchanged not allowing for any energy gains. Further,
it is noted that in the limit of $\beta_1 \to 0$ and $\beta_{\mathrm{BH}} \to 0$,
no mass gain should be possible and indeed this is apparent since the
solution becomes $m_3 = m_1 \sqrt{1 - 4 \kappa_{i1}^2}$ i.e. $m_3 < m_1$ for
all $\kappa_{i1}>0$.

\subsection{Equilibrium velocity}

Consider starting at rest, $\beta_1=0$, and emitting a photon which
gives a final velocity such that the spaceship ends up with a maximally
increased mass. This can be calculated by taking the limit of 
Equation~(\ref{eqn:masschange}) for $\beta_1 \to 0$ and then differentiating
$\partial[\lim_{\beta_1\to0} m_3]/\partial \kappa_{i1} = 0$ solving
for $\kappa_{i1}$. This occurs when

\begin{align}
\kappa_{i1} &= \frac{1}{2} \Big( \frac{ \beta_{\mathrm{BH}} }{ 1 + \beta_{\mathrm{BH}} } \Big),
\end{align}

giving a final mass of

\begin{align}
\frac{m_3}{m_1} &= \gamma_{\mathrm{BH}},
\end{align}

where $\gamma_{\mathrm{BH}} = (1-\beta_{\mathrm{BH}}^2)^{-1/2}$. Evaluating the
corresponding velocity, which is labelled as the ``equilibrium velocity'' in what
follows ($\beta_{\mathrm{eq}}$):

\begin{align}
\beta_{\mathrm{eq}} = \beta_{\mathrm{BH}},
\end{align}

which has an intuitive interpretation since at parity speed $\nu_f = \nu_i$.

\subsection{Terminal velocity}
\label{sub:terminal}

This gained mass can now be used to induce further acceleration. Whilst
this could be achieved by simply exhausting photons, the most efficient
means would be again to use the BH mirror and exploit the halo drive.

Let us take Equation~(\ref{eqn:masschange}), set $\beta_1\to\beta_{\mathrm{BH}}$
since the starting speed is the equilibrium speed, and solve the expression
to be equal to $1/\gamma_{\mathrm{BH}}$ with respect to $\kappa_{i1}$. The
photon energy needed is easily found to be given by 

\begin{align}
\kappa_{i1} &= \frac{\beta_{\mathrm{BH}}}{2} \sqrt{\frac{1-\beta_{\mathrm{BH}}}{1+\beta_{\mathrm{BH}}}},
\end{align}

and plugging into our $\beta_3$ equation where $\beta_1$ is to again initiated
from $\beta_{\mathrm{BH}}$, one obtains a ``terminal velocity'',
$\beta_{\mathrm{term}}$, of

\begin{align}
\beta_{\mathrm{term}} &= \frac{ 2 \beta_{\mathrm{BH}} }{ 1 + \beta_{\mathrm{BH}}^2 },
\label{eqn:termvel}
\end{align}

which is bound to be $0\leq\beta_{\mathrm{term}}<1$ for all
$0\leq\beta_{\mathrm{BH}}<1$, as expected. Expanding to third-order in
$\beta_{\mathrm{BH}}$, $\beta_{\mathrm{term}}$ may be written as

\begin{align}
\beta_{\mathrm{term}} &= 2 \beta_{\mathrm{BH}} - 2 \beta_{\mathrm{BH}}^3 + \mathcal{O}[\beta_{\mathrm{BH}}^5].
\label{eqn:termvelexp}
\end{align}

To first order then, the terminal velocity equals twice that of the black hole,
consistent with the first-order result for a conventional gravitational
slingshot. In essence then, one is conducting a remote slingshot using the halo
rather than physically approaching the BH and risk tidal disruption (as well as
an increased flight time and heavy time dilation by diving into the
gravitational well).

From \citet{kipping:2017}, one expects the following two statement to be true, if
the principle of ensemble equivalence holds. First, rather than accelerating
from rest to equilibrium speed with one photon, and then equilibrium to terminal
with a second photon, the same acceleration could be achieved for the same energy
using a large number of smaller photon energies. This point is important because
it is impractical to emit such a high energy photon in a single step. Second,
if this principle holds, then the reverse should also be true and both steps
should be achievable in a single photon i.e. one should be able to accelerate
from rest to terminal with a single photon emission.

This latter statement may be verified by solving $\lim_{\beta_1 \to 0} m_3 = m_1$
with respect to $\kappa_{i1}$ - the single high energy photon, which yields
a quadratic solution of

\begin{equation}
\kappa_{i1} =
\begin{cases}
0,\\
\frac{\beta_{\mathrm{BH}}}{1 + \beta_{\mathrm{BH}}}\\
\end{cases}.
\end{equation}

The zero result clearly corresponds to no motion at all.
Plugging the latter result into our $\beta_3$ equation in the limit where
$\beta_1 \to 0$ yields the same terminal velocity as that stated in
Equation~(\ref{eqn:termvel}), in accordance with the principle.

It's worth comparing this single photon emission to that of the maximum
photon emission earlier, $\kappa_{i1,\mathrm{max}}$. One may easily show
that $\lim_{\beta_1 \to 0} \kappa_{i1,\mathrm{max}}$ equals this single photon
energy if, and only if, $\beta_{\mathrm{BH}} = 1$. This therefore
re-enforces that this physical limit cannot be practically achieved.

\subsection{Accounting for relativistic Doppler shifts}
\label{sub:doppler}

One important effect thus far ignored is the relativistic Doppler
shift of the returning photon in the spacecraft's frame of motion.
Even for a single photon emission, the emission causes a back-reaction
which accelerates the spacecraft away from rest up to $\beta_2$.
Accordingly, when the photon returns it is not reabsorbed as $\nu_f$
but as $\nu_f''$, where the dashes indicate a Lorentz transform to
the rest frame of the spacecraft.

Following the principle of photon equivalence, one can simplify the
derivation by considering a single photon emission to accelerate
up to terminal velocity - defined as the maximum speed for
which $m_3 = m_1$. Starting from rest, $\beta_2$ and $m_2$ are
the same as Equations~(\ref{eqn:beta2}) \& (\ref{eqn:mass2}) found
earlier, except that $\beta_1 \to 0$, giving

\begin{align}
\lim_{\beta_1 \to 0} \beta_2 &= \frac{\kappa_{i1}}{1 - \kappa_{i1}},\\
\lim_{\beta_1 \to 0} \frac{m_2}{m_1} &= \sqrt{1 - 2 \kappa_{i1}}.
\end{align}

Before, it was assumed that the photon returned with a frequency given by
Equation~(\ref{eqn:infmass}) in the limit of $\alpha\to\pi$. One may now
modify this to

\begin{align}
\nu_f &= \nu_i \Big( \frac{ 1 + \beta_{\mathrm{BH}} }{ 1 - \beta_{\mathrm{BH}}} \Big)
\sqrt{\frac{1 - \beta_2}{1 + \beta_2}},
\label{eqn:nufcorr}
\end{align}

where the square root term accounts for the relativistic Doppler shift.
When this photon returns, the final mass, $m_3$, can be
calculated using Equation~(\ref{eqn:mass3}) except the photon energy
is substituted using Equation~(\ref{eqn:nufcorr}), yielding

\begin{align}
\frac{m_3}{m_1} &= (1 - 2 \kappa_{i1}) \Bigg( 1+ 2 \kappa_{i1} \sqrt{1-2 \kappa_{i1}} \Big( \frac{1+\beta_{\mathrm{BH}}}{1-\beta_{\mathrm{BH}}}\Big) \Bigg).
\end{align}

Solving $m_3 = m_1$ with respect to $\kappa_{i1}$ yields

\begin{align}
\kappa_{i1} &= \frac{1}{2} \Bigg( 1 - \Big(\frac{1-\beta_{\mathrm{BH}}}{1+\beta_{\mathrm{BH}}}\Big)^{2/3} \Bigg).
\label{eqn:critri1corr}
\end{align}

Plugging this result into Equation~(\ref{eqn:beta3}) yields a revised terminal
velocity (after much simplification) of

\begin{align}
\beta_{\mathrm{term}} &= \frac{ (1+\beta_{\mathrm{BH}})^{4/3} - (1-\beta_{\mathrm{BH}})^{4/3} }{
(1+\beta_{\mathrm{BH}})^{4/3} + (1-\beta_{\mathrm{BH}})^{4/3}
},
\label{eqn:termvelcorr}
\end{align}

which is again bound to be $0\leq\beta_{\mathrm{term}}<1$ for all
$0\leq\beta_{\mathrm{BH}}<1$. Expanding to third-order for small
$\beta_{\mathrm{BH}}$, $\beta_{\mathrm{term}}$ may be written as

\begin{align}
\beta_{\mathrm{term}} &= \frac{4}{3} \beta_{\mathrm{BH}} - \frac{28}{81} \beta_{\mathrm{BH}}^3 + \mathcal{O}[\beta_{\mathrm{BH}}^5].
\label{eqn:termvelcorrexp}
\end{align}

In the limit of large $\beta_{\mathrm{BH}}$, Equation~(\ref{eqn:termvelcorr})
is well-approximated by $\gamma_{\mathrm{term}} \simeq 2^{1/3}
\gamma_{\mathrm{BH}}^{4/3}$. These results show that the Doppler shifts
decrease the amount of energy transferred to the spacecraft, but nevertheless
speeds in excess of the black hole's velocity can be achieved.

\subsection{Accounting for gravitational red/blue-shifts}
\label{sub:GRshift2}

As discussed earlier in Section~\ref{sub:GRshift1}, gravitational
red/blue shifts can be shown to be an extremely small effect so long
as $M c^2 \gg h \nu_i$ and the photon returns to the same radial
distance. According to Equation~(\ref{eqn:ri1max}), $\kappa_{i1}<\tfrac{1}{2}$
and thus $m c^2 < h \nu_i/2$. Accordingly, the valid regime can also
be stated as $M \gg m$. However, since the objective of the halo drive is
to accelerate the spacecraft to relativistic velocities, then clearly
the geodesic will be chosen such that the photon does not in fact return
to the same location but rather a greater radial distance.

Consider starting from rest and attempting to accelerate to terminal
velocity with a single photon of energy given by
Equation~(\ref{eqn:critri1corr}). The intermediate velocity of the
spacecraft is $\beta_2$, which here can be evaluated to be

\begin{align}
\beta_2 =& \frac{\kappa_{i1}}{1 + \kappa_{i1}},\nonumber\\
\qquad&= \frac{ 1 - \big( \frac{ 1 - \beta_{\mathrm{BH}} }{ 1 + \beta_{\mathrm{BH}} } \big)^{2/3} }{ 1 + \big( \frac{ 1 - \beta_{\mathrm{BH}} }{ 1 + \beta_{\mathrm{BH}} } \big)^{2/3} }.
\end{align}

The time interval for the photon to return is approximately
$(2d_0 + \Delta d)/c$ and thus the distance traversed is

\begin{align}
\Delta d &\simeq \frac{2 d_0 \beta_2}{1 - \beta_2}.
\end{align}

The gravitational redshift from $d_0$ to $d_0 + \Delta d$, when the
photon returns, is given by

\begin{align}
\nu_{f,\mathrm{corr}} &\simeq \nu_f \sqrt{ \frac{(d_0+\Delta d)(d_0 - R_S)}{ d_0 (d_0 + \Delta d - R_S) } }.
\end{align}

This effect can be considered to be insignificant if the relativistic Doppler
correction made in the previous subsection far exceeds the change caused
by the gravitational red shift, i.e. when

\begin{align}
1 - \sqrt{ \frac{(d_0 + \tfrac{2 d_0 \beta_2}{1 - \beta_2} )(d - R_S)}{ d (d + \tfrac{2 d_0 \beta_2}{1 - \beta_2} - R_S) } } \ll 1 - \sqrt{ \frac{1 - \beta_2}{1 + \beta_2} }.
\end{align}

\begin{figure}
\begin{center}
\includegraphics[width=8.4cm,angle=0,clip=true]{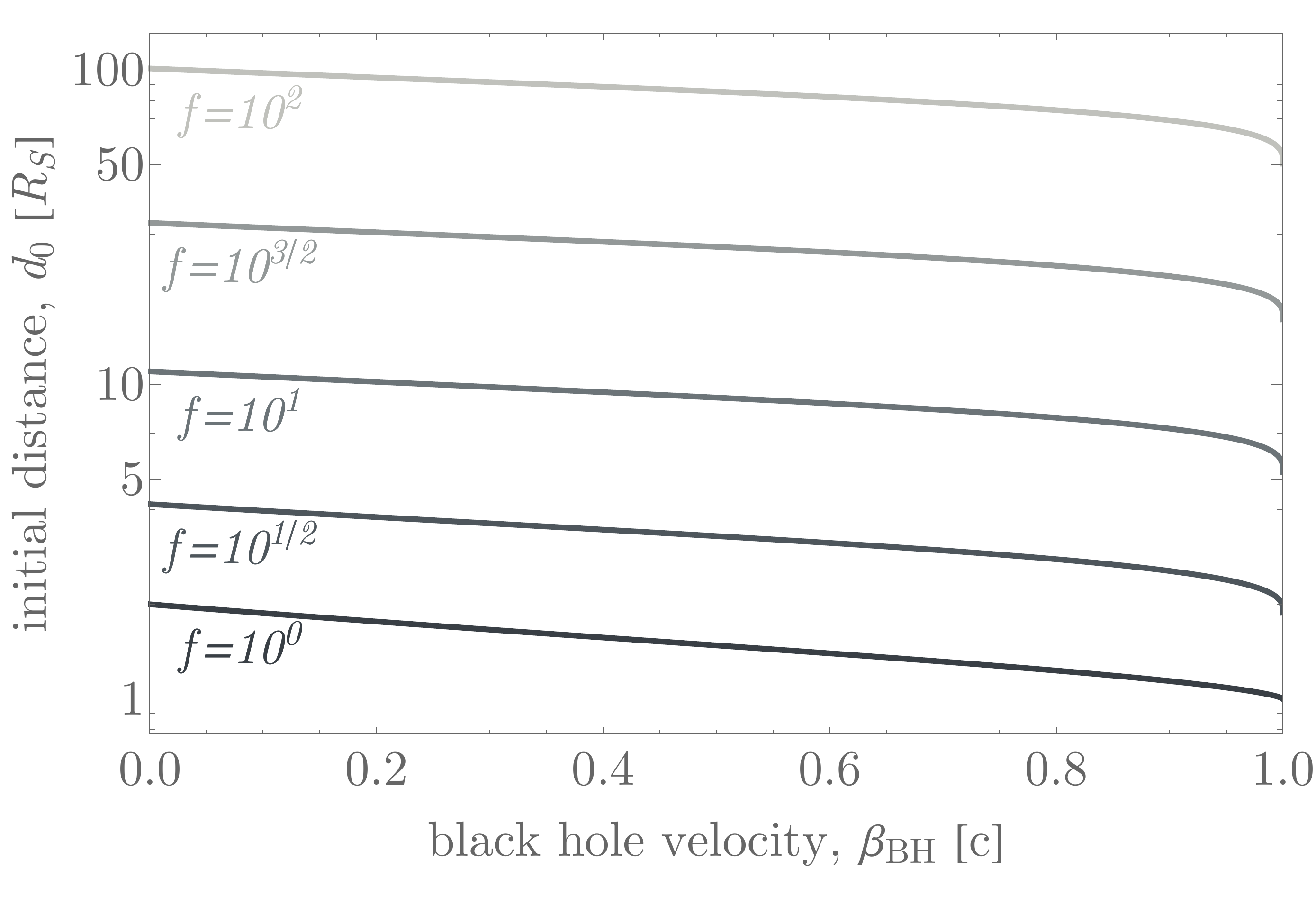}
\caption{
Initial stand-off distance from the black hole such that gravitational
redshift effects are $f$ times smaller than those of Doppler effects
when using the halo drive. The approximation $d_0 \gg R_S$ found
analytically in the limit of $\beta_{\mathrm{BH}} \to 0$ generally
holds up well except for extreme cases.
}
\label{fig:gravredshift}
\end{center}
\end{figure}

Solving for $d_0$ in the limit of $\beta_1 \to 0$, one can show that equates to
the condition that $d \gg R_S$, where $R_S$ is the Schwarzschild radius. If the
ratio between the RHS and the LHS of the above is labelled as $f$, then
Figure~\ref{fig:gravredshift} demonstrates that this argument works well
even for relatively high $\beta_{\mathrm{BH}}$. Accordingly, it is argued that the
terminal velocity derived in Equation~(\ref{eqn:termvelcorrexp}) is accurate
so long as $d_0 \gg R_S$, in which case additional effects such as changes in
the relative binary position leading to time-dependent gravitational redshifts
can also be safely ignored.

\subsection{Numerical tests}

Throughout this work, it has been assumed that the principle of ensemble equivalence
described in \citet{kipping:2017} also holds here, although this has not
been tested. The problem closely resembles that described in
\citet{kipping:2017} and thus generally it is expected to hold. Further,
Section~\ref{sub:terminal} showed that a single-photon acceleration
produced the same results as that of the double-photon acceleration curve.
Of course, a single photon emitted with an energy comparable to the rest mass
of the spacecraft is not feasible (or indeed desirable) and generally
implementation would involve the emission of a large sequence of lower energy
photons to produce a more gradual acceleration.

It is therefore worthwhile to test whether the terminal velocity predicted from
a single photon model indeed equals that when a large number of sequential
emissions are performed instead. Using the equations described throughout
this work, a calculation was performed for the acceleration for $N$ photons of equal
frequencies set to $\nu = (m c^2 \kappa_{i1})/(h N)$, where $\kappa_{i1}$ is
set to the value derived earlier necessary to achieve terminal velocity in the
case of a single photon. If the principle holds, then the final velocity
after numerically integrating $N$ sequential steps should equal the
terminal velocity (to within floating point precision).

\begin{figure}
\begin{center}
\includegraphics[width=8.4cm,angle=0,clip=true]{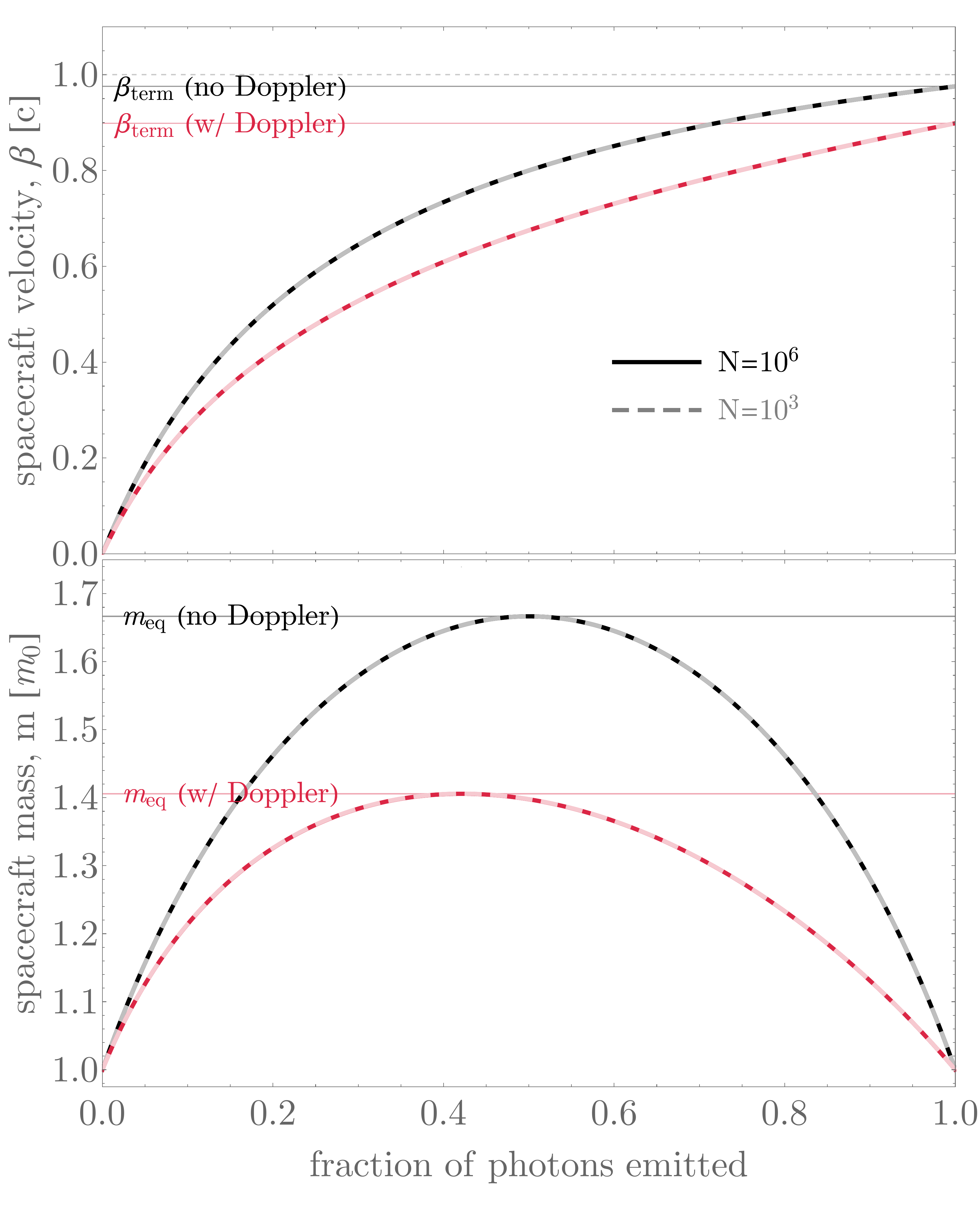}
\caption{
Numerical tests of the principle of ensemble equivalence for
$\beta_{\mathrm{BH}}=0.8$. As expected, photons can be released either in a
small number at high energies or in a large number of equivalent cumulative
energy, but the results are the same. The dashed lines show the predictions
from our earlier derivations in the case of a single photon assumption.
Accounting for Doppler shifts leads to significant changes in the results.
}
\label{fig:numerical}
\end{center}
\end{figure}

As shown in Figure~\ref{fig:numerical}, it is easy to verify that the
principle holds and more over it is possible to accurately predict the
terminal velocity of the spacecraft using our formulae.

\section{Discussion}
\label{sec:discussion}

\subsection{Response of the binary \& observational signatures}
\label{sub:evolution}

The halo drive causes a spacecraft of essentially arbitrary mass (so long as
$m \ll M_{\odot}$) to accelerate up to relativistic speeds (for suitably
compact binaries) without losing any fuel in the process\footnote{Note
that solutions do exist for moving $m \sim M_{\odot}$ via an alternative
mechanism, as described in \citet{shkadov:1987,forgan:2013}.}. At face value,
this makes remote detection of halo drives seemingly impossible. But, there is
no such thing a free lunch and of course something here has lost energy and
that's the binary itself. By the time it reaches terminal velocity, the
spacecraft has increased its energy from $m c^2$ to $m c^2
\gamma_{\mathrm{term}}$, and accordingly one can write that the binary
must have lost an energy of

\begin{align}
\Delta E &= (\gamma_{\mathrm{term}} - 1) m c^2,\nonumber\\
\qquad&= \frac{m c^2}{2} \Bigg( \frac{ \big( (1+\beta_{\mathrm{BH}})^{2/3} - (1-\beta_{\mathrm{BH}})^{2/3} \big)^2 }{ (1-\beta_{\mathrm{BH}}^2)^{2/3} } \Bigg).
\label{eqn:DeltaE}
\end{align}

The binding energy of a binary system is given by

\begin{align}
E &= -\frac{G M M_2}{2 a} + \mathcal{O}[\tfrac{1}{c^2}],
\end{align}

where $a$ is the binary separation and $M_2$ is the mass of the secondary
component. If the binary evolves from $a$ to $a - \Delta a$ as a result 
of the kick, then one may show that to first-order in $\Delta E/E$

\begin{align}
\frac{\Delta a}{a} &= -\frac{2 a \Delta E}{G M M_2}.
\end{align}

If one writes that $a = \tilde{a} G M/c^2$ (i.e. in half Schwarzschild radii), then

\begin{align}
\frac{\Delta a}{a} &= - \tilde{a} \frac{m}{M_2} \Bigg( \frac{ \big( (1+\beta_{\mathrm{BH}})^{2/3} - (1-\beta_{\mathrm{BH}})^{2/3} \big)^2 }{ (1-\beta_{\mathrm{BH}}^2)^{2/3} } \Bigg).
\end{align}

Expanding the $\Delta E$ term to second-order in $\beta_{\mathrm{BH}}$,
accurate to 3\% for all $\beta_{\mathrm{BH}}<0.5$, yields

\begin{align}
\frac{\Delta a}{a} &\simeq - f \frac{m}{M_2} \beta_{\mathrm{BH}}^2.
\label{eqn:neworbit}
\end{align}

This reveals that the binary will be kicked into a slightly
eccentric orbit with the periapsis position located at the
extraction point, with the new semi-major axis shrinking by
that described by Equation~(\ref{eqn:neworbit}). In general, these
changes are small for all $m \ll M_2$ or effectively all
$m \ll M_{\odot}$.

A civilization using a network of binaries may not only accelerate
from them but also decelerate upon return, thus potentially undoing the
slight distortions made to the binary. Even so, the binary temporarily
spends time at closer semi-major axis where gravitational radiation
is more effective and thus one still expects elevated merger rates to
result.

One-way trips, perhaps from a central hub, would lead to an even higher
rate of binary in-spiral on-top of the natural gravitational radiation.
If journeys are made isotropically, an eccentric binary may not
result but accelerated in-spiral would persist. However, only a discrete
set of highways exist between galactic binary black holes and thus the
distortions can never be perfectly isotropic meaning that excess
eccentricity would likely persist.

\subsection{From infinitesimal to finite beams}

One effect ignored in the earlier derivation is that it was assumed that the
beam has an infinitesimal width. In reality, the beam has a finite width
and that width will diverge in a physically real system. It is therefore
critical that the beam divergence over the entire path length is less than the
size of the spacecraft's receiver, $L$, else significant energy losses would
occur.

Beams will diverge due to two effects. The first of these is via diffraction,
and for a diffraction limited beam one expects the width to diverge
after a distance $2d$ to

\begin{align}
W_r = W_t + \frac{2\sqrt{2}d\lambda}{D_t},
\end{align}

where $D_t$ is the diameter of the transmitter and $W$ denotes the
width of the beam at reception and transmission. If the spacecraft has
a physical width of $L$, then one requires

\begin{align}
\lambda \ll \frac{ L D_t c^2}{ 2\sqrt{2} (d/R_S) 2 G M },
\end{align}

or

\begin{align}
\lambda \ll 120\,\mu\mathrm{m}\,\Big(\frac{L}{D_t}\Big) \Big(\frac{D_T}{10\,\mathrm{m}}\Big)^2 \Big(\frac{d}{100 R_S}\Big)^{-1}
\Big(\frac{M_{\odot}}{M}\Big).
\label{eqn:diffraction}
\end{align}

In the neighborhood of the black hole, within a hundred Schwarzschild
radii, it should be easy to produce collimated electromagnetic radiation at
such wavelengths. This is can be extended to much greater distances
if the receiver is much larger than the transmitter ($L \gg D_T$).
This latter point is particularly relevant because the halo drive is able
to accelerate effectively arbitrarily large masses up to $\beta_{\mathrm{term}}$
(so long as $m \ll M_{\odot}$) allowing for extremely large (e.g. planet-sized)
vehicles. Ultimately, diffraction can be overcome by simply using shorter
wavelength light, or even particle beams. For this reason,
although diffraction is an unavoidable effect, it could be mitigated against
unless halo drives are attempted at extreme distances where it may become
impractical to emit/absorb such high energy radiation.

A second effect that leads to beam divergence comes from essentially a tidal effect.
Consider a beam which has finite width and is emitted at a single angle,
$\delta$, tailored such that the center of the beam will perform a boomerang
geodesic (e.g. using the method described in Section~\ref{sub:geodesic}). Photons
emitted slightly off to the side beam's center will encounter the black hole
at slightly different impact parameters. Since the beam angle is chosen such that
only the center line performs a boomerang, then the edges will saddle the separatrix
and experience distinct deflection angles, leading to the effect of achromatic beam
divergence.

Let's say that the edge of the beam is offset from the center by a distance $W_t/2$.
A light ray emitted from this point crosses the radial line between the
black hole and the center of the beam at a distance $d + W_t/(2\tan \delta)$.
Accordingly, the correct angle this photon should be emitted at to perform
a boomerang is not $\delta$, but rather (using the result from
Figure~\ref{fig:haloexamples}):

\begin{align}
\delta_{\mathrm{edge}} &= \frac{\delta_0}{d + \tfrac{W_t}{2}\cot\tfrac{\delta_0}{d}}.
\end{align}

Accordingly, the beam would potentially miss the spacecraft upon return. The
key problem is that the photons at the edge of the beam were emitted at the
wrong angle, $\delta$, whereas the correct boomerang angle would have been
$\delta_{\mathrm{edge}}$.

This point reveals that the problem actually stems from the way in
which the beam was chosen to be setup - a planar source such that the entire
beam has the same initial emission angle. For this reason, the
divergence is not unavoidable in the same sense as diffraction is, but rather
is primarily an engineering problem that could be surmountable through
careful beam shaping (see \citealt{dickey:2003}). The purpose of this work is
not to provide an actual blueprint for the halo drive, but rather merely
highlight that no physical barrier exists to prevent such a scheme.
Nevertheless, one possible solution could be a large number of micro-emitters
with independent actuators that would be combined to form the overall beam,
where each micro-emitter has a unique angular displacement to correct for the
effect, analogous to how adaptive optics corrects wavefront errors in the
Earth's atmosphere using individual actuators. Clearly, such a system would
require a very advanced control system to make the necessary calculations for
each actuator, but again there's no obvious physical barrier to overcoming this
problem.

\subsection{Ignored effects}

It is important to highlight several approximations made in this work. The purpose of this
paper is to introduce the concept of using halos as described, and thus
several small effects were ignored to facilitate the calculations that are
briefly discussed here.

First, this work has assumed that extremely efficient absorption of the photon
is assumed by the spacecraft upon reception. An idealized system needs to be
able to recycle the photons with thermal losses (see \citealt{slovick:2013})
much smaller than the total energy transferred to the spacecraft, $\Delta E$.

A second effect ignored is the energy to overcome the gravitational potential energy of
the binary in order to escape the system. Tacitly, it was assumed that the
velocities achieved far exceed the escape velocity from the initial standoff
distance. Requiring $\Delta E$ of Equation~(\ref{eqn:DeltaE}) to be much
greater than the gravitational potential energy of a binary where $M_2 = q M$,
one may show that

\begin{align}
\frac{d}{R_S} \gg& \Bigg(\frac{1+q}{2}\Bigg) \Bigg( \frac{ (1-\beta_{\mathrm{BH}}^2)^{2/3} }{ \big( (1+\beta_{\mathrm{BH}})^{2/3} - (1-\beta_{\mathrm{BH}})^{2/3} \big)^2 } \Bigg),\nonumber\\
\frac{d}{R_S} \gg& \Bigg( \frac{27 \beta_{\mathrm{BH}}^{-2} - 22 }{48} \Bigg),
\end{align}

where on the second line, right hand bracket has been Taylor expanded to first-order
as well as assuming  $q \sim 1$. For low $\beta_{\mathrm{BH}}$, such
as $\beta_{\mathrm{BH}}=0.05$, this requires a large stand-off distance of
a couple of thousand Schwarzschild radii. In the mildly relativistic scenario
of $\beta_{\mathrm{BH}}=0.2$, standoff distances greater than around a hundred
Schwarzschild radii would make the gravitational potential energy factor
much smaller than the gained energy. Nevertheless, it could be worthwhile
to include this generally small contribution in future work.

A third assumption is that the circumbinary environment
is devoid of opaque material that would lead to beam losses. For example,
an accretion disk around the black hole would certainly make it a
sub-optimal target for a halo drive. Accordingly, if one requires compact
binaries for relativistic acceleration, the other component would need
to be another black hole or neutron star to avoid mass transfers forming
a disk.

\subsection{Other applications of the halos}
\label{sub:otherapps}

Numerous earlier works have highlighted the potential use of black
holes for advanced technological applications (e.g. see
\citealt{crane:2009,inoue:2011}) and the halo drive provides another example.

Although not the focus of this work, it is worth highlighting that halo drives
could have other purposes besides from just accelerating spacecraft.
For example, the back reaction on the black hole taps energy from it,
essentially mining the gravitational binding energy of the binary. Similarly,
forward reactions could be used to not only decelerate incoming spacecraft
but effectively store energy in the binary like a fly-wheel, turning
the binary into a cosmic battery.

Another possibility is that the halos could be used to deliberately manipulate
black holes into specific configurations, analogous to optical tweezers. This
could be particularly effective if halo bridges are established between
nearby pairs of binaries, causing one binary to excite the other. Such cases
could lead to rapid transformation of binary orbits, including the deliberate
liberation of a binary.

It is also highlighted that acceleration could be performed in a two-body process where
the source is a very massive emitter in the system but the halo strikes a
second nearby and lower mass vehicle. This vehicle could then be accelerated
to even faster velocities than the terminal velocity computed earlier.
Such a system would lead to the more massive source also experiencing
a kick back into a higher orbit, as well transferring some fraction of its
initial mass to the accelerated vehicle. Thus, the system would have a
finite lifetime before the accelerator would reach very large orbital
radii where halos would become difficult to establish via diffraction
constraint of Equation~(\ref{eqn:diffraction}).

\subsection{Kerr metrics}
\label{sub:penrose}

This work has focused on halo drives being applied to a Schwarzschild black
hole \citep{schwarzschild:1916} in a compact binary system. However, it is
hypothesized here that lone, isolated Kerr black holes \citep{kerr:1963} could
likely serve the same function. By riding along the frame dragged spacetime
surrounding the black hole, light should be blue shifted (in the case of same
sense revolution), permitting the rotational energy of the black to be
tapped\footnote{It is highlighted that \citet{cramer:1997} calculate boomerang
geodesics for Kerr black holes but the blue shift effect was not considered.}.
This joins the numerous ways previously proposed to extract energy from Kerr
black holes, such as the Penrose process \citep{penrose:1971}, superradiance
with amplifying incident waves for various fields \citep{zeldovich:1972,
bardeen:1972,starobinsky:1973,churilov:1973,teukolsky:1974} and the
Blandford-Znajek process \citep{blandford:1997}. Calculation of the Kerr-case
was beyond the scope of this work but would be an interesting problem for
the future.

\section{Conclusions}
\label{sec:conclusions}

The search for intelligence amongst the cosmos is often guided by considering
the possible activities of hypothetical advanced civilizations and the
associated technosignatures that would result (e.g. \citealt{dyson:1960,
lin:2014,korpela:2015}). At the same time, there is growing interest in
developing the means for humanity to take our first steps into becoming an
interstellar civilization (e.g. \textit{Breakthrough Starshot}; see
\citealt{parkin:2018}). These two enterprises can often overlap, since
advanced propulsion systems may lead to observable technosignatures
(e.g. \citealt{guillochon:2015}). Along these lines, this work has
considered how an advanced civilization might utilize the light sailing
concept to conduct relativistic and extremely efficient propulsion.

The proposed system is that a spacecraft emits a collimated beam of energy
towards at a black hole at a carefully selected angle, such that the beam
returns to the spacecraft - a so-called boomerang geodesic
\citep{stuckey:1993}. If the black hole is moving towards the spacecraft,
as could be easily accomplished by exploiting a compact binary, this halo
of particles will return with a higher energy (and momentum). This energy is
then transferred to the spacecraft allowing for acceleration. Overall then,
the halo drive transfers kinetic energy from the moving black hole to the
spacecraft by way of a gravitational assist.

The analysis presented assumes the halo is photonic, but the beam could be
comprised of massive particles too and achieve the same effect. Either way,
the system described echoes the \citet{dyson:1963} slingshot, except that
the spacecraft does not physically slingshot around the compact object,
but rather let's the light beam do the slingshot on its behalf.

An appealing aspect of the halo drive is that no fuel is spent. The
spacecraft gradually gains energy during its initial acceleration and
then discharges that energy for further acceleration up to terminal
velocity - the speed at which the spacecraft returns to its original
mass.

The terminal velocity of the spacecraft is 133\% the black hole's speed,
to first-order. Critically, this velocity in not sensitive to the mass of
the spacecraft, with the only assumption being that said mass is much less
than that of the black hole. Accordingly, a major advantage of the halo
drive is that Jupiter-mass spacecraft could be accelerated to relativistic
speeds.

Beam divergence due to tidal effects on a finite beam width could be
mitigated by careful beam shaping. Divergence due to diffraction is not
expected to lead to noticeable losses for large spacecraft using
optical lasers within a hundred Schwarzschild radii. Nevertheless,
for this reason, the system is argued to be impractical at distances much
greater than this, thereby necessitating relatively expedient acceleration.

An advanced civilization utilizing such a system would first have to
have achieved interstellar flight to journey towards the nearest suitable
BH. They could then could use BHs in binary systems as way-points throughout
the galaxy, of which there are likely $\mathcal{O}[10^7]$ in the Milky Way
\citep{reggiani:2013}, serving as both acceleration and deceleration stations.
Alternatively, they could use the larger population of
BHs which do not reside in compact binaries \citep{elbert:2017} via their
proper motions, although this would not permit for such high velocities.

Each departure from a binary in a particular direction kicks the binary into
a slightly eccentric orbit and accelerates
it's in-spiral merger rate. In principle, each arrival from the same
direction would undo this effect leading to no observable signature.
However, finite time differences between the departure and arrival
would cause the binary to spend time at a tighter semi-major axis
than it would naturally, during which time it would experience more
rapid gravitational radiation in-spiral. Accordingly, a possible
technosignature of the halo drive would be an enhanced rate of black
hole binary in-spiral, versus say their neutron star counterparts.

\section{Acknowledgements}

DMK is supported by the Alfred P. Sloan Foundation.
Thanks to Nick Stone, Zephyr Penoyre, Zoltan Haiman, Jerry Ostriker,
Janna Levin, Avi Loeb and Claes Cramer for helpful conversations in preparing
this manuscript. I would also like to thank Bill Stuckey for his correspondence
regarding gravitational mirrors, and Michael Hippke and Duncan Forgan
for constructive comments on an early draft of this paper. I am also grateful
to the anonymous reviewers for their helpful comments.

%% This command is needed to show the entire author+affilation list when
%% the collaboration and author truncation commands are used.  It has to
%% go at the end of the manuscript.
%\allauthors

%% Include this line if you are using the \added, \replaced, \deleted
%% commands to see a summary list of all changes at the end of the article.
%\listofchanges

\end{document}